\documentclass{aa}
\usepackage{graphics,epsfig,txfonts}
\usepackage{multirow}

\usepackage{natbib}
\bibpunct{(}{)}{;}{a}{}{,}

\newcommand{\oexpo}[1]{$10^{#1}$}

\newcommand{\ltsima}{$\buildrel < \over \sim$}
\newcommand{\lsim}{\lower.5ex\hbox{\ltsima}}
\newcommand{\gtsima}{$\buildrel > \over \sim$}
\newcommand{\gsim}{\lower.5ex\hbox{\gtsima}}
\newcommand{\msun}{M$_{\odot}$}
\newcommand{\xmm}{XMM-Newton}

\def\rahour{\hbox{\ensuremath{^{\rm h}}}}
\def\ramin{\hbox{\ensuremath{^{\rm m}}}}

\newcommand{\xmmp}{\hbox{IGR\,J05414-6858}}

\usepackage{color}

\begin{document}
 
\title{Discovery of the neutron star spin and a possible orbital period from the Be/X-ray binary IGR J05414-6858 in the LMC.
       \thanks{Based on observations with 
               XMM-Newton, an ESA Science Mission with instruments and contributions 
               directly funded by ESA Member states and the USA (NASA)}
      }

\author{      R.~Sturm\inst{1} 
     \and     F.~Haberl\inst{1}  
     \and     A.~Rau\inst{1} 
     \and     E. S. Bartlett\inst{2}
     \and     X.-L.~Zhang\inst{1}
     \and     P.~Schady\inst{1}
     \and     W.~Pietsch\inst{1} 
     \and     J.~Greiner\inst{1}
     \and     M. J. Coe\inst{2} 
     \and     A.~Udalski\inst{3}  
       }

\titlerunning{The Be/X-ray binary IGR J05414-6858.}
\authorrunning{Sturm et al.}

\institute{  Max-Planck-Institut f\"ur extraterrestrische Physik, Giessenbachstra{\ss}e, 85748 Garching, Germany
	    \and
            School of Physics and Astronomy, University of Southampton, Highfield, Southampton SO17 1BJ, United Kingdom
	    \and
            Warsaw University Observatory, Aleje Ujazdowskie 4, 00-478 Warsaw, Poland
	   }

\date{Received 22 February 2012 / Accepted 29 April 2012}

 \abstract{The number of known Be/X-ray binaries in the Large Magellanic Cloud is small compared to the observed population of the Galaxy or the Small Magellanic Cloud.
           The discovery of a system in outburst provides the rare opportunity to measure its X-ray properties in detail.}
          {\xmmp\ was discovered in 2010 by INTEGRAL and found in another outburst with the Swift satellite in 2011.
           In order to characterise the system, we analysed the data from a follow-up \xmm\ target of opportunity observation of the 2011 outburst
           and investigate the stellar counterpart with photometry and spectroscopy.}
          {We modelled the X-ray spectra from the EPIC instruments on \xmm\ and compared them with Swift archival data. 
           In the X-ray and optical light curves, we searched for periodicities and variability.
           The optical counterpart was classified using spectroscopy obtained with ESO's Faint Object Spectrograph at NTT.}
          {The X-ray spectra as seen in 2011 are relatively hard with a photon index of $\sim$0.3--0.4 and show only low absorption. 
           They deviate significantly from earlier spectra of a probable type II outburst in 2010.
           The neutron star spin period of $P_{\rm spin} = 4.4208$ s was discovered with EPIC-pn.
           The $I$-band light curve revealed a transition from a low to a high state around MJD 54500.
           The optical counterpart is classified to B0-1~IIIe and shows H$\alpha$ emission and a variable NIR excess, vanishing during the 2010 outburst.
           In the optical high state, we found a periodicity at 19.9 days, probably caused by binarity and indicating the orbital period.}
           {}

\keywords{galaxies: individual: Large Magellanic Cloud --
          stars: emission-line, Be -- 
          stars: neutron --
          X-rays: binaries}
 
\maketitle

\section{Introduction}
\label{sec:introduction}

\begin{table*}
 \caption{X-ray observations of \xmmp.}
 \begin{center}
   \begin{tabular}{lcccccrrrr}
     \hline\hline\noalign{\smallskip}
     \multicolumn{1}{c}{Observation} &
     \multicolumn{1}{c}{ObsID} &
     \multicolumn{1}{c}{Date} &
     \multicolumn{1}{c}{Time} &
     \multicolumn{1}{c}{Instrument} &
     \multicolumn{1}{c}{Mode\tablefootmark{a}} &
     \multicolumn{1}{c}{Offax\tablefootmark{b}} &
     \multicolumn{1}{l}{Net Exp} &  
     \multicolumn{1}{r}{Net Cts\tablefootmark{c}} &
     \multicolumn{1}{r}{R$_{\rm sc}$\tablefootmark{d}}\\

     \multicolumn{1}{l}{} &
     \multicolumn{1}{l}{} &
     \multicolumn{1}{l}{} &
     \multicolumn{1}{c}{(UT)} &
     \multicolumn{1}{l}{} &
     \multicolumn{1}{l}{} &
     \multicolumn{1}{c}{[\arcmin]} &
     \multicolumn{1}{c}{[ks]} &   
     \multicolumn{1}{l}{} &
     \multicolumn{1}{r}{[\arcsec]}\\
     \noalign{\smallskip}\hline\noalign{\smallskip}

          XMM 2001  &  0094410101  &  2001-10-19  &  17:49 -- 20:30  &  EPIC-pn    &  ff--medium  &  8.2  &  8.2    & $<$26  &  30  \\
                    &              &              &  17:09 -- 20:29  &  EPIC-MOS1  &  ff--medium  &  7.2  &  11.7   & $<$5   &  30  \\
                    &              &              &  17:09 -- 20:29  &  EPIC-MOS2  &  ff--medium  &  6.9  &  11.7   & $<$14  &  30  \\

     \noalign{\smallskip}\hline\noalign{\smallskip}

          XMM 2011  &  0679380101  &  2011-08-13  &  07:45 -- 14:07  &  EPIC-pn    &  ff--thin    &  1.1  &  7.2   & 2018  &  21  \\
                    &              &              &  07:22 -- 14:07  &  EPIC-MOS1  &  ff--medium  &  1.2  &  9.4   &  863  &  29  \\
                    &              &              &  07:22 -- 14:07  &  EPIC-MOS2  &  ff--medium  &  1.2  &  9.4   &  954  &  31  \\

     \noalign{\smallskip}\hline\noalign{\smallskip}

          Swift 2010 a &  00031745001  &  2010-06-25  &  05:35 -- 09:10  &  XRT    &  pc    &  5.3   &  3.9   & 221   &  40  \\
          Swift 2010 b &  00031745002  &  2010-06-30  &  01:14 -- 23:59  &  XRT    &  pc    &  2.0   &  5.2   & 497   &  50  \\

     \noalign{\smallskip}\hline\noalign{\smallskip}

         Swift 2011 a &  00045428001  &  2011-08-05  & 15:28 -- 22:26   &  XRT    &  pc    &  10.5   &  3.3  &  40  &  40  \\
         Swift 2011 b &  00031745003  &  2011-08-09  & 23:53 -- 04:55   &  XRT    &  pc    &  1.8   &  3.7  & 120  &  40  \\
         Swift 2011 c &  00031745004  &  2011-08-12  & 00:03 -- 22:53   &  XRT    &  pc    &  2.4   &  4.1  & 100  &  40  \\

         Swift 2011 d &  00031745005  &  2011-08-20  & 00:25 -- 16:53   &  XRT    &  pc    &  3.1   &  2.1  &  $<$6  &  40  \\
         Swift 2011 e &  00031745006  &  2011-08-24  & 02:44 -- 04:29   &  XRT    &  pc    &  2.5   &  1.2  &  $<$5  &  40  \\

      \noalign{\smallskip}\hline\noalign{\smallskip}
    \end{tabular}
 \end{center}
  \tablefoot{
  \tablefoottext{a}{Observation setup: full-frame mode (ff) and photon counting mode (pc). For \xmm, also the filter is given.}
  \tablefoottext{b}{Off-axis angle under which the source was observed.}
  \tablefoottext{c}{Net counts as used for spectral analysis in the (0.2 -- 10.0) keV band for \xmm\ and in the (0.3 -- 6.0) keV band for {\em Swift}.}
  \tablefoottext{d}{Radius of the circular source extraction region.}
  }
 \label{tab:xray-obs}
\end{table*}

Depending on the donor star, high mass X-ray binaries (HMXBs) are divided into super-giant systems and Be/X-ray binaries \citep[BeXRBs, e.g. ][]{2011Ap&SS.332....1R}.
In the latter case, a not well understood mechanism causes matter ejection of the Be star in the equatorial plane, leading to the build up of an equatorial decretion disc around the Be star \citep{2001PASJ...53..119O}.
This disc dominates the emission of the system in infrared and some emission lines like H$\alpha$. The variability of this emission points to the instability of such discs.
Due to a supernova kick, a neutron star (NS) can have an eccentric orbit around the Be star. During periastron passage the NS can accrete matter from the decretion disc, causing a so-called type I X-ray outburst, 
enduring several days at typical luminosities of $10^{36}$ erg s$^{-1}$. During disc instabilities, the NS can accrete a large fraction of the decretion disc, resulting in type II outbursts with luminosities up to $10^{37}$ erg s$^{-1}$ for several weeks.

The transient behaviour of Be/X-ray binaries and the wide extent of the Large Magellanic Cloud (LMC) on the sky, 
which imposes a low observational coverage by X-ray missions, 
complicate the discovery and investigation of Be/X-ray binaries in this galaxy. 
In contrast to that, the SMC was monitored with RXTE for about 14 years \citep{2008ApJS..177..189G}.
Therefore, only nine HMXB X-ray pulsars are known to date in the LMC, which inhibits a statistical comparison of this sample with those of the Galaxy and the Small Magellanic Cloud (SMC). 
In the Galaxy and the SMC $\sim$66 and $\sim$55 HMXB pulsars are known, respectively.
A major fraction of the pulsars is found in Be/X-ray binaries \citep[e.g. ][]{2010ASPC..422..224C}.
Population studies of these systems are important to understand the stellar evolution, as they e.g.
allow to estimate supernova kick velocities \citep{2005MNRAS.358.1379C} or the star formation history \citep{2010ApJ...716L.140A,2011AN....332..349M}.
Recently, a bimodal NS spin period distribution for the Galactic and SMC samples
has been associated with two different types of supernovae \citep{2011Natur.479..372K}.
To enable such statistical studies for the LMC, it is necessary to successively build up a larger sample of X-ray measurements of pulsars in outburst.

In 2010, \xmmp\ was discovered serendipitously within INTEGRAL observations of SN\,1987A \citep{2010ATel.2695....1G} and later localised \citep{2010ATel.2696....1L} and 
identified as Be-X-ray binary \citep{2010ATel.2704....1R} with {\em Swift} and GROND follow-up observations. 
In 2011, {\em Swift} performed a UV survey of the LMC (PI: S. Immler). 
This provided a shallow coverage of this galaxy with the {\em Swift} X-ray telescope (XRT) and allowed the detection of bright X-ray transients.
In an observation on 2011 Aug. 5 an outburst of \xmmp\ was detected \citep{2011ATel.3537....1S}, 
which allowed us to request an \xmm\ target of opportunity (ToO) observation.

In this study, we report our analysis of the \xmm\ observation of \xmmp. 
The detection of the NS spin period adds the tenth X-ray pulsar in the LMC sample 
and a detailed X-ray spectral and temporal analysis allows a characterisation of the system.
We compare our new results to those from archival {\em Swift} data and discuss complementary optical data to characterise the optical counterpart and the circum-stellar disc.

\section{Observations and data reduction}
\label{sec:observations}

\subsection{XMM-Newton}
The \xmm\ \citep{2001A&A...365L...1J} ToO observation was performed on 2011 Aug. 13. 
The source was observed on-axis, placed on CCD4 of EPIC-pn \citep{2001A&A...365L..18S} and CCD1 of both EPIC-MOS \citep{2001A&A...365L..27T} detectors.
We used \xmm\ SAS 11.0.0\footnote{Science Analysis Software (SAS), http://xmm.vilspa.esa.es/sas/} to process the data. 
Unfortunately, the observation was affected by an increased background caused by soft protons.
During the first $\sim$11.5 ks of the $\sim$17 ks observation, the background was at a moderately elevated level, 
allowing the selection of time intervals where the background rate in the (7.0--15.0) keV band was below 50 cts ks$^{-1}$ arcmin$^{-2}$ for EPIC-pn and below 4 cts ks$^{-1}$ arcmin$^{-2}$ for EPIC-MOS.
The detailed observation setup is recorded in Table~\ref{tab:xray-obs}. 
Here, we also list an \xmm\ observation from 2001 covering the position of \xmmp, in which the source was not detected. We used this observation to derive an upper limit for the flux.
For EPIC-pn, we used single and double pixel events and single to quadruple events in the case of EPIC-MOS, all having {\tt FLAG=0}.
Background events were selected from a point source free area on the same CCDs as the source.
Source events were extracted from a circle, with radius optimised for the signal-to-noise ratio by the SAS task {\tt eregionanalyse}.
We created spectra and response matrices with {\tt especget} and used a binning to have at least a signal to noise ratio of 5 for each bin.
For time series, the photon arrival times were randomised within the CCD frame time and calculated for the solar system barycentre.

\subsection{Swift}
We re-analysed archival {\em Swift}/XRT observations.
The spectra were created by using the {\tt ftool}\footnote{http://heasarc.nasa.gov/ftools/} {\tt xselect} 
to select events in the cleaned level 3 event files within a circle, placed on the source with radii given in Table~\ref{tab:xray-obs}.
Background spectra were created from a circular extraction region with radius of 200\arcsec.
The spectra were binned to have $\geq$20 cts bin$^{-1}$. The ancillary response files were calculated with {\tt xrtmkarf}.
{\em Swift} observations of \xmmp\, including non-detections, are also listed in Table~\ref{tab:xray-obs}.

\subsection{NIR, optical, and UV photometry}

Optical photometry of IGR~J05414--6858 was obtained with OGLE, GROND and {\em Swift}/UVOT.
%
%
The optical counterpart was monitored regularly during the Optical Gravitational Lensing Experiment (OGLE) of phase III \citep[][]{2008AcA....58...69U} between October 2001 and April 2009 in the $I$-band.
The source identification is OGLEIII\,LMC175.4.21714.

The Gamma-ray Burst Optical Near-ir Detector  
\citep[GROND;][]{2008PASP..120..405G}   at   the  MPG/ESO   2.2m
telescope in La Silla, Chile, observed the source at three epochs in June 2010 and January
2012.   Preliminary  results of  the  2010  observations were  already
presented in \citet{2010ATel.2704....1R}.
GROND is  a 7-channel imager that  observes in four  optical and three
near-IR  channels  simultaneously.   The  IGR~J05414--6858  data  were
reduced and analysed with the  standard tools and methods described in
\cite{2008ApJ...685..376K}.     The   photometry    was    obtained   using
point-spread-function   (PSF)   fitting   taking   into   account   the
contamination  from  the  two  nearby sources  (see Fig.~\ref{fig:fc}).
Calibration  was performed  against observations  of an  SDSS standard
star  field  ($g^{\prime}r^{\prime}i^{\prime}z^{\prime}$)  or  against
selected  2MASS  stars  \citep{2006AJ....131.1163S}  ($J H  K_S$).   This
resulted in  1$\sigma$ accuracies of  0.04\,mag ($g^\prime z^\prime$),
0.03\,mag  ($r^\prime  i^\prime$),  0.05\,mag  ($JH$),  and  0.07\,mag
($K_S$).

The Ultraviolet/Optical Telescope (UVOT) aboard {\em Swift} has three optical ($v$ $b$ $u$) and three UV filters ($uvw1$ $uvm2$ $uvw2$).
\xmmp\ was observed in all UVOT filters during the pointed X-ray observations.
For the LMC UV-survey observation 00045428001, the source is not in the field of view of the UVOT.
Photometry was carried out on pipeline processed sky images downloaded from the {\em Swift} data centre\footnote{http://www.swift.ac.uk/swift\_portal}, 
following the standard UVOT procedure \citep{2008MNRAS.383..627P}.


\subsection{Optical spectroscopy} 
Optical spectroscopy was taken with the ESO Faint Object Spectrograph (EFOSC2) mounted at the Nasmyth B focus of the 3.6m New Technology Telescope (NTT), La Silla, Chile on the nights of 2011 December 8 and 10. 
The EFOSC2 detector (CCD\#40) is a  Loral/Lesser, Thinned, AR coated, UV flooded, MPP chip with 2048$\times$2048 pixels corresponding to 4.1\arcmin$\times$4.1\arcmin on the sky. 
The instrument was in longslit mode with a slit width of 1.5\arcsec. 
Grisms 14 and 20 were used for blue and red end spectroscopy respectively. 
Grism 14 has a wavelength range of $\lambda\lambda3095$--$5085$~\AA{} and a grating of 600~lines~mm$^{-1}$ and a dispersion of 1~\AA{}~pixel$^{-1}$. 
The resulting spectra have a spectral resolution of  $\sim12$~\AA{}. 
Grism 20 is one of the two new Volume-Phase Holographic grisms recently added to EFOSC2. 
It has a smaller wavelength range, from 6047--7147~\AA{}, but a superior dispersion of 0.55 \AA{}~pixel$^{-1}$ and 1070 lines~mm$^{-1}$. 
This produced a spectral resolution for our red end spectra of $\sim$6~\AA{}. 
Filter OG530 was used to block second order effects.
The data were reduced using the standard packages available in the Image Reduction and Analysis Facility \textsf{IRAF}. Wavelength calibration was implemented using comparison spectra of Helium and Argon lamps taken throughout the observing run with the same instrument configuration. The spectra were normalized to remove the continuum and a redshift correction applied corresponding to the recession velocity of the LMC \citep[-280~km~s$^{-1}$, ][]{2002LEDA.........0P}.

\section{Analyses and results of X-ray data}
\label{sec:analyses}

\subsection{X-ray coordinates}
\label{sec:analyses:coord}

\begin{figure}
  \resizebox{\hsize}{!}{\includegraphics[angle=0,clip=]{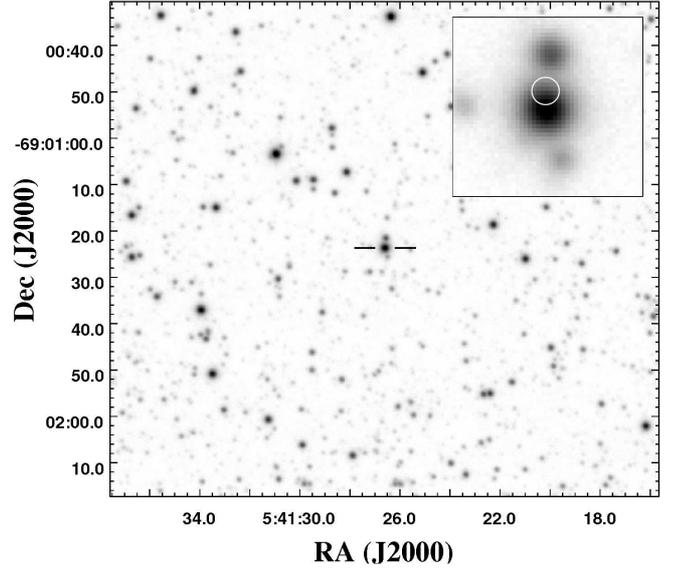}}
  \caption{ GROND $r'$-band finding chart. Lines mark the counterpart of \xmmp. 
            In the zoom-in, the \xmm\ position is marked by a white circle with radius of the 1$\sigma$ position uncertainty of 0.52\arcsec.
            The astrometric solution accuracy of the GROND image is 0.47\arcsec\ in RA and 0.21\arcsec\ in Dec.
          }
  \label{fig:fc}
\end{figure}

We created X-ray images from all three EPIC cameras in the \xmm\ standard energy sub-bands.
A simultaneous source detection was performed on these images with {\tt edetect\_chain}.
The best-fit source position is 
RA (J2000) = 05\rahour41\ramin26\fs62 and 
Dec (J2000) = $-$69\degr01\arcmin23\farcs0.
The $1\sigma$ uncertainty of the position is 0.52\arcsec, where we assume a systematic error of 0.5\arcsec, which is quadratically added to the statistical error.
The angular separation to the optical counterpart is 0.52\arcsec\ for the 2MASS position
and 0.69\arcsec\ for the GROND position \citep[star A, ][]{2010ATel.2704....1R}.
The distance to the {\em Swift} position of \citet[][]{2010ATel.2696....1L} is 2.0\arcsec\ with an uncertainty in the {\em Swift} measurement of $\sim$3\arcsec. 
A finding chart obtained from GROND data is presented in Fig.~\ref{fig:fc}. The white circle in the zoom-in gives the \xmm\ position.
The improved X-ray coordinates from the \xmm\ observation further confirm the identification of the X-ray source with the optical counterpart.

\subsection{Spectral analysis}
\label{sec:analyses:spec}

Spectral analysis was performed with {\tt xspec} \citep{1996ASPC..101...17A} version 12.7.0.
The three \xmm\ EPIC spectra were fitted simultaneously and we included constant factors in the models to consider instrumental differences.
For all models, we obtained consistent values of $C_{\rm MOS1}=1.11\pm0.08$ and $C_{\rm MOS2} = 1.10\pm0.08$  relative to EPIC-pn ($C_{\rm pn}=1$).
Therefore, the fluxes for all instruments are consistent within uncertainties, 
as EPIC-MOS is known to derive $\sim$5\% higher values compared to EPIC-pn.
All other model parameters were forced to be the same for all instruments.
The spectra are well described by an absorbed power-law.
The photoelectric absorption was modelled by a fixed Galactic foreground column density of $N_{\rm H, Gal} = 6\times 10^{20}$ cm$^{-2}$ \citep{1990ARA&A..28..215D} 
with abundances according to \citet{2000ApJ...542..914W}. 
An additional column density $N_{\rm H, LMC}$ was determined in the fit.
It accounts for the interstellar medium of the LMC and source intrinsic absorption
and the abundances were set to 0.5 for elements heavier than helium \citep{1992ApJ...384..508R}.

\begin{figure}
  \resizebox{\hsize}{!}{\includegraphics[angle=-90,clip=]{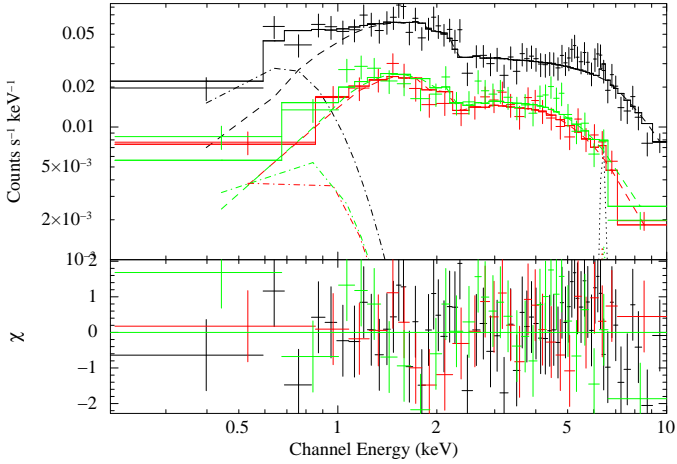}}
  \caption{ EPIC-pn (black), EPIC-MOS1 (red), EPIC-MOS2 (green) spectra, together with the best-fit po+bb+Fe model (solid line) and its individual components: power-law (dashed), black-body (dashed-dotted) and Fe line (dotted).
            The lower panel shows the residuals. 
          }
  \label{fig:spec_xmm}
\end{figure}

We further tested the spectra for the existence of typical features of BeXRBs.
A possible soft excess \citep[e.g.][]{2008A&A...491..841E,2004ApJ...614..881H} was modelled by a black-body (Model: PL+BB) or a multi-temperature disc black-body model (PL+DiscBB).
The additional model component only improves the fit marginally, but demonstrates model-dependent uncertainties for the power-law parameters. 
The f-test probability for these additional components is around 13\%.
We use the black-body model to derive an upper limit for a soft component contribution.
A soft excess can account only for $<$1\% of the detected flux and up to 22.0\% of the unabsorbed luminosity in the (0.2--10.0)~keV band.
The derived radius of the emission region from the black body is too large for an NS. 
With the disc model we obtain a similar inner radius for an inclination of $\Theta=0$ ($R_{\rm in} \propto (\cos\Theta)^{-1/2}$).
Also according to \citet{2004ApJ...614..881H}, we obtain an inner disc radius of $R_{\rm in} = ( L_{\rm X} / 4\pi\sigma T^4)^{1/2} = 52$ km. 
Fluorescent iron line emission at 6.4 keV was modelled with a Gaussian line, having a fixed central energy and no measurable broadening.
Also here, we receive only a marginal improvement of the fit and a line flux of $2.8\pm2.6 \times 10^{-6}$ photons cm$^{-2}$ s$^{-1}$.
The spectrum and best-fit model are presented in Fig.\ref{fig:spec_xmm} and the best-fit results are listed in Table~\ref{tab:spectra}.
All uncertainties and limits correspond to a 90\% confidence level ($\Delta \chi^2 = 2.71$).

\begin{figure}
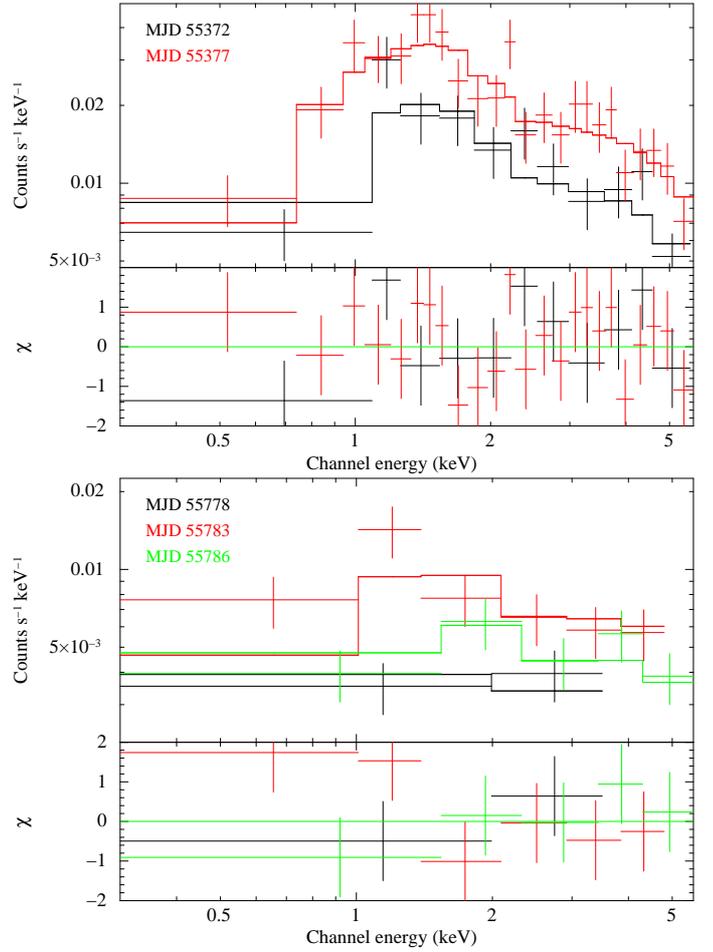

  \resizebox{\hsize}{!}{\includegraphics[angle=-90,clip=]{spec_swift2010.ps}}
  \resizebox{\hsize}{!}{\includegraphics[angle=-90,clip=]{spec_swift2011.ps}}
  \caption{ {\em Swift} spectra of \xmmp\ from 2010 ({\it top}) and  2011 ({\it bottom}) with best-fit power-law model. Lower panels give the residuals. 
          }
  \label{fig:spec_swift}
\end{figure}

To investigate the spectral variability, we compared the EPIC spectra to the archival {\em Swift} spectra. 
The lower statistics only allowed to fit the power-law model. 
For the two spectra in 2010, $N_{\rm H, LMC}$ and $\Gamma$ are consistent within errors.
Also, no significant evolution is found in the three {\em Swift} spectra of 2011.
We  assumed that the spectral shape is time-independent during the individual outbursts,
and fitted the model simultaneously to the two 2010 and three 2011 {\em Swift} spectra (see Fig.~\ref{fig:spec_swift}).
The fluxes were determined for all observations individually.
The results are also listed in Table~\ref{tab:spectra}.
There is significant spectral variability observed between the outbursts of \xmmp\ in 2010 and 2011,
where the recent outburst exhibits a harder X-ray spectrum, consistent with the \xmm\ 2011 observation.
If we fit the {\em Swift} 2010 data with the spectral model derived from the \xmm\ 2011 observation (allowing a re-normalisation),
the fit degrades from $\chi^2_{\rm red} = 1.03$ to 1.75.
In contrast to that, the $\chi^2_{\rm red}$ improves from 1.18 to 1.06 ($\chi^2=10.6$, dof$=10$), 
if we fit the Swift 2011 spectra with the \xmm-derived model.
This is caused by the increase of the degrees of freedom. 

Since the lower photon index in 2011 is seen with \xmm\ and {\em Swift}, 
this is unlikely caused by the high background during the \xmm\ observation or by instrumental differences.

\begin{table*}
  \caption[]{Spectral fit results.}
  \begin{center}
    \begin{tabular}{rlccccccccc}
      \hline\hline\noalign{\smallskip}
      \multicolumn{1}{l}{Observation} &
      \multicolumn{1}{l}{Model\tablefootmark{a}} &
      \multicolumn{1}{c}{$N_{\rm H, LMC}$} &
      \multicolumn{1}{c}{$\Gamma$} &
      \multicolumn{1}{c}{$kT$} &
      \multicolumn{1}{c}{$R$\tablefootmark{b}} &
      \multicolumn{1}{c}{EW$_{\rm Fe}$} &
      \multicolumn{1}{c}{Flux\tablefootmark{c}} &
      \multicolumn{1}{c}{$L_{\rm x}$\tablefootmark{d}} &
      \multicolumn{1}{c}{$\chi^2_{\rm red}$} &
      \multicolumn{1}{c}{dof} \\
      \multicolumn{1}{c}{} &
      \multicolumn{1}{c}{} &
      \multicolumn{1}{c}{[\oexpo{21}cm$^{-2}$]} &
      \multicolumn{1}{c}{} &
      \multicolumn{1}{c}{[eV]} &
      \multicolumn{1}{c}{[km]} &
      \multicolumn{1}{c}{[eV]} &
      \multicolumn{1}{c}{[$10^{-12}$ erg cm$^{-2}$ s$^{-1}$]} &
      \multicolumn{1}{c}{[$10^{36}$ erg s$^{-1}$]} &
      \multicolumn{1}{c}{} &
      \multicolumn{1}{c}{} \\
      \noalign{\smallskip}\hline\noalign{\smallskip} \vspace{.7mm}
      XMM 2011  &  PL            & $<$0.89           & $0.32_{-0.05}^{+0.03}$ & --             & --             &  --            & $3.01_{-0.32}^{+0.24}$ & 0.91 &  0.92 & 118 \\ \vspace{.7mm}
                &  PL+BB         & $4.98_{-3.8}^{+3.4}$ & $0.40_{-0.09}^{+0.07}$ & $112_{-23}^{+56}$ & $62_{-46}^{+66}$ &  --            & $2.98_{-0.35}^{+0.20}$ & 0.98 &  0.91 & 116 \\ \vspace{.7mm}
                &  PL+BB+Fe      & $4.98_{-3.6}^{+3.6}$ & $0.41_{-0.09}^{+0.07}$ & $111_{-23}^{+49}$ & $62_{-45}^{+73}$ &  $78_{-70}^{+71}$ & $2.97_{-0.47}^{+0.37}$ & 0.98 &  0.88 & 115 \\ 
                &  PL+DiscBB     & $3.68_{-1.4}^{+2.6}$ & $0.39_{-0.08}^{+0.08}$ & $128_{-33}^{+85}$ & $49_{-44}^{+134}$ &  --            &  $2.98_{-0.33}^{+0.16}$ & 1.01 &  0.91 & 116 \\

      \noalign{\smallskip}\hline\noalign{\smallskip}\vspace{.7mm}

      Swift 2010 a  &  \multirow{2}{*}{PL}        & \multirow{2}{*}{$1.9_{-1.4}^{+2.0}$}  & \multirow{2}{*}{$0.78_{-0.17}^{+0.18}$} &  \multirow{2}{*}{--}             & \multirow{2}{*}{--}             &  \multirow{2}{*}{--}            & 
                                                                                             $5.6_{-1.6}^{+1.1}$ & 1.77 &  \multirow{2}{*}{1.03} & \multirow{2}{*}{31} \\ \vspace{.7mm}
      Swift 2010 b  &            &                   &                     &                &                &                & $9.2_{-1.9}^{+1.3}$ & 2.91 &       &    \\ 

      \noalign{\smallskip}\hline\noalign{\smallskip}\vspace{.7mm}

      Swift 2011 a &             &            &                     &                &                &                & $5.4_{-5.4}^{+2.6}$ & 1.62 &       &     \\ \vspace{.7mm}
      Swift 2011 b &  PL         &   $<$2.6   & $0.25_{-0.22}^{+0.14}$ &  --            &   --           &      --        & $5.1_{-4.5}^{+1.3}$ & 1.51 &  1.18  & 8    \\ 
      Swift 2011 c &             &            &                     &                &                &                & $3.5_{-3.1}^{+9.8}$ & 1.04 &        &     \\ 

      \noalign{\smallskip}\hline
    \end{tabular}
  \end{center}
  \tablefoot{
\tablefoottext{a}{For definition of spectral models see text.}
\tablefoottext{b}{Radius of the emitting area (for BB) or inner disc radius for an inclination of $\Theta=0$ (for DiscBB, for the definition see text).}
\tablefoottext{c}{Observed flux in the (0.2--10.0) keV band, derived by integrating the best-fit model.}
\tablefoottext{d}{Source intrinsic X-ray luminosity in the (0.2--10.0) keV band corrected for absorption and assuming a distance of the source of 50 kpc.}
  }
  \label{tab:spectra}
\end{table*}

\subsection{Pulsations}
\label{sec:analyses:pulsation}

\begin{figure}
  \resizebox{\hsize}{!}{\includegraphics[angle=-90,clip=]{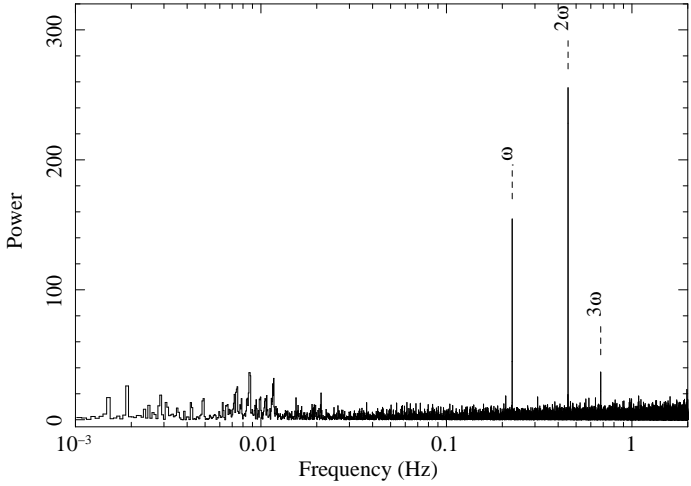}}
  \caption{
           Power density spectrum of \xmmp\ for the EPIC-pn time series in the (0.2--10.0) keV band. 
           The best-fit frequency  of $\omega =  0.2262$ Hz and its first and second harmonics are marked with dashed lines.
          }
  \label{fig:pds}
\end{figure}

\begin{figure}
  \resizebox{\hsize}{!}{\includegraphics[angle=0,clip=]{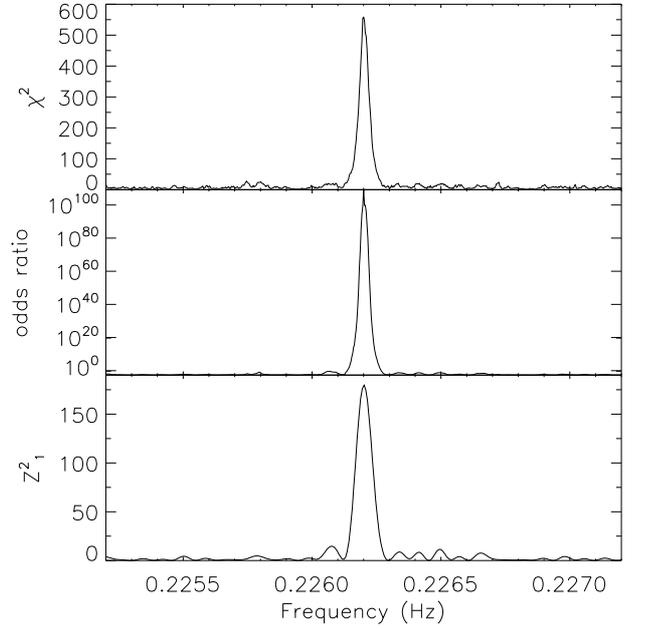}}
  \caption{
           {\it Top:} $\chi^2$ test for persistence of the EPIC-pn light curve, around trial frequencies between 0.2252 and 0.2272 Hz.
           {\it Middle:} Frequency dependence of the Bayesian odds ratio.
           {\it Bottom:} Rayleigh $Z^2_1$ test.
          }
  \label{fig:period_sta}
\end{figure}

A strong signal  at $\omega = 0.2262$ Hz and its first harmonic appeared in a Fast Fourier Transformation (FFT) of the EPIC-pn time series in the (0.2--10.0) keV band.
The power density spectrum is plotted in Fig.~\ref{fig:pds}.
The signal is also clearly present in the (0.2--2.0) keV and (2.0--10.0) keV sub-bands.
The period is not resolved by EPIC-MOS (2.6 s frame time), 
since the period is shorter than twice the frame time of the instrument, i.e. $\omega$  is above the Nyquist frequency. 
The same holds for the {\em Swift} data (2.5 s frame time).
A $\chi^2$ test, a Bayesian odds ratio \citep{1996ApJ...473.1059G}, 
and a Rayleigh Z$^2$ test for one harmonic \citep{2002A&A...391..571H,1983A&A...128..245B} 
around the periodicity signal are shown in Fig.~\ref{fig:period_sta}.
All tests independently confirm the pulse period.
Following \citet{2008A&A...489..327H}, we used the Bayesian detection method 
to determine the pulse period and a 1$\sigma$ uncertainty of 4.420866(2) s on 2011-08-13.

\begin{figure}
  \resizebox{\hsize}{!}{\includegraphics[angle=0,clip=]{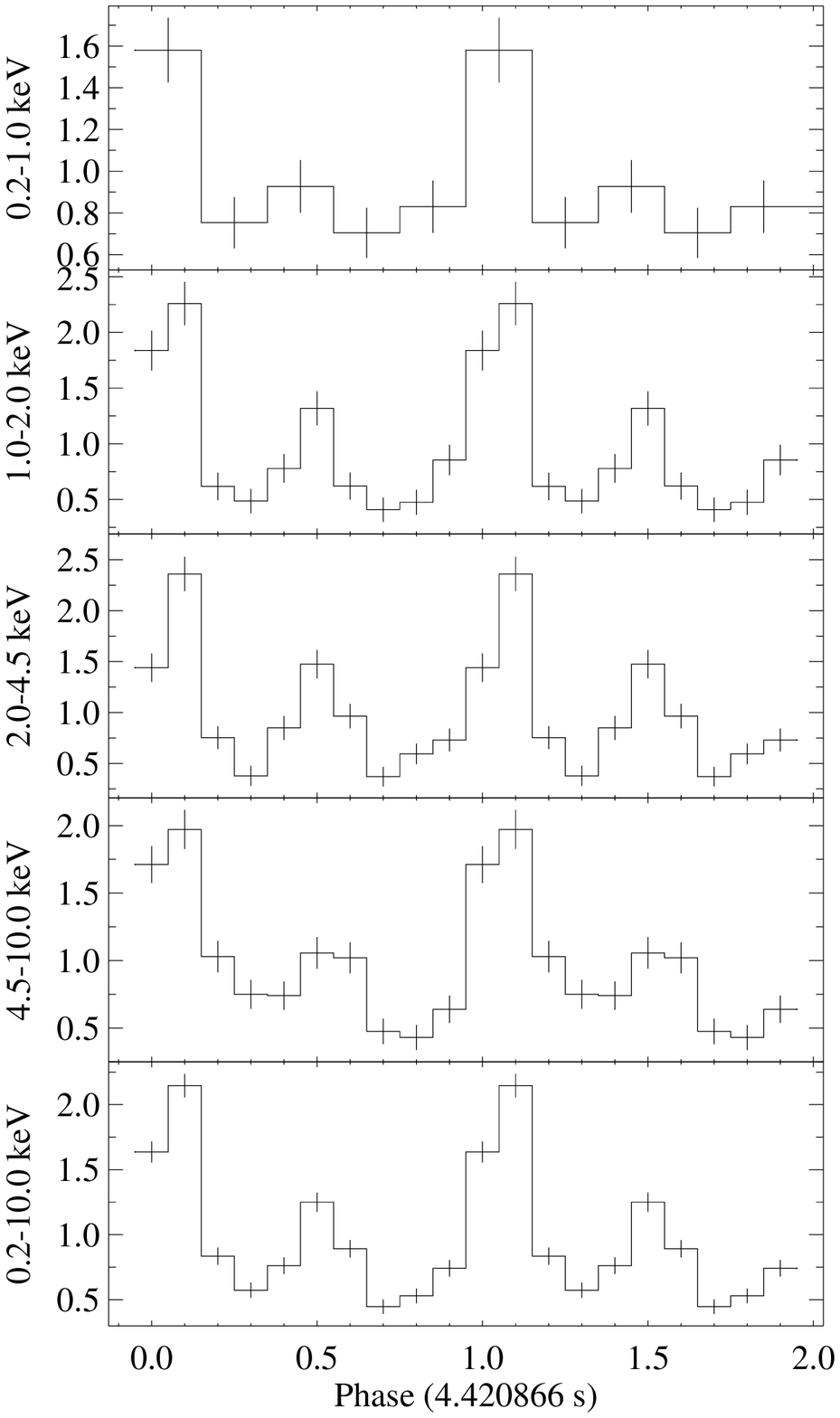}\includegraphics[angle=0,clip=]{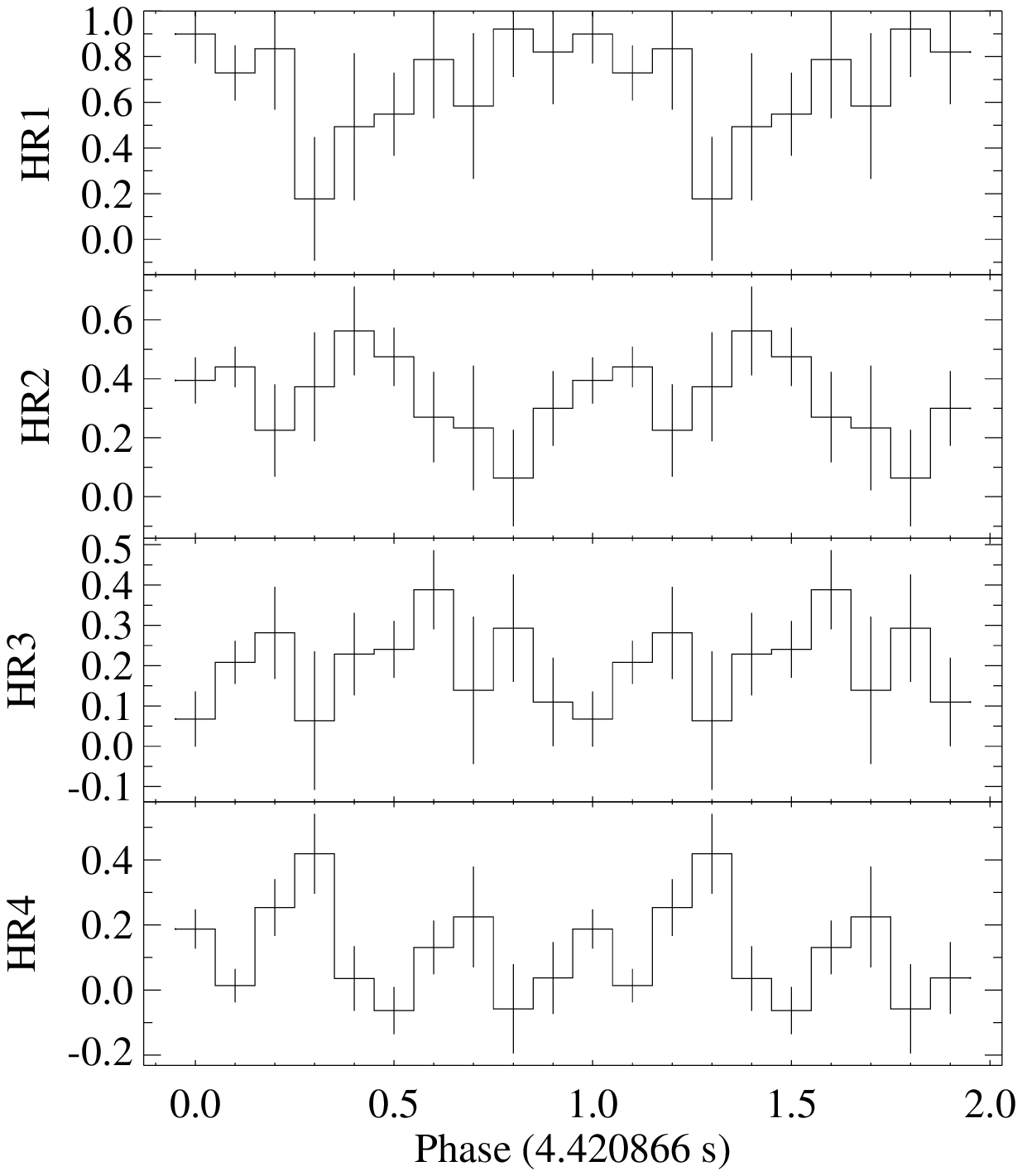}} 
  \caption{{\it Left:} X-ray pulse profile of \xmmp\ in various energy bands from the EPIC-pn time series.
           The pulse profiles are background subtracted and normalised to the average net count rate of 
           3.0, 4.6, 6.8, 8.3 and 22.4 $\times 10^{-2}$ cts s$^{-1}$ from top to bottom.
           {\it Right:} Hardness ratios as a function of pulse phase derived from the pulse profiles in two neighbouring standard energy bands.
          }
  \label{fig:pp}
\end{figure}

Fig.~\ref{fig:pp} shows the folded background-subtracted light curves from EPIC-pn in the total (0.2 -- 10.0) keV band 
and the standard sub-bands (0.2 -- 0.5) keV, (0.5 -- 1.0) keV, (1.0 -- 2.0) keV, (2.0 -- 4.5) keV, and (4.5 -- 10.0) keV,
where we merged the first two bands, to increase the statistics.
Also hardness-ratio (HR) variations are presented.
HRs are defined by HR$_{i}$ = (R$_{i+1}$ $-$ R$_{i}$)/(R$_{i+1}$ + R$_{i}$) with R$_{i}$ denoting the background-subtracted 
count rate in the standard energy band $i$ (with $i$ from 1 to 4).
In the light curves, two narrow peaks are seen within one period having only small variations in energy.
By modelling the (0.2 -- 10.0) keV curve with a non-pulsating contribution and one Gaussian for each peak, 
we estimate a pulsed fraction of (48$\pm$7)\% in the total flux, and a flux ratio for both peaks of 2.3$\pm$0.5.

A search for pulsations in the INTEGRAL ISGRI observations of 2010 was performed.
We detect the source at $0.28\pm0.04$ cts s$^{-1}$ in the (20--40) keV band, 
but could not find a significant period in the power density spectrum or variability in the 4.4208~s folded light curve.
This might be caused by binary orbital modulations, as the INTEGRAL observations cover a long time.

\subsection{X-ray flux variability}
\label{sec:analyses:time}

In addition to the fluxes as measured in Sec.~\ref{sec:analyses:spec},
we calculated upper limits for non-detections of \xmmp.
The field was observed by \xmm\ in 2001, but no source was detected by {\tt edetect\_chain} (analogous to Sec.~\ref{sec:analyses:coord}).
Spectra were extracted in the same manner as described in Sec.~\ref{sec:observations}
from a 30\arcsec\ source region and a 50\arcsec\ background region.
Using C statistics and the spectral shape as determined with \xmm\ in 2011,
we derived a 90\% upper limit for the flux of 9.0$\times 10^{-14}$ erg cm$^{-2}$ s$^{-1}$.
This lowest upper limit results in a variability of at least a factor of 100,
compared to the maximum flux measured in the {\em Swift} 2010b observation.
Prior to that, we could not find any corresponding X-ray detection.
There is no ROSAT source within 1\arcmin\ listed in the literature.

The {\em Swift} monitoring of the recent outburst, determined the turn-off between MJD 55786 and 55793.
For the two non-detections, we extracted spectra in the same way as above and determined 90\% confidence upper limits
of $9.5\times 10^{-13}$ erg cm$^{-2}$ s$^{-1}$ and $2.3\times 10^{-12}$ erg cm$^{-2}$ s$^{-1}$.
The long term evolution of the system is presented in Fig.~\ref{fig:lc}.

The variability during the \xmm\ observation in 2011 was at a moderate level.
We created a  background corrected light curve, merged using all EPIC instruments and binned to 70 s, corresponding to 30 cts bin$^{-1}$ on average.
A $\chi^2$ test of this light curve against a constant resulted in  $\chi^2/{\rm dof} = 130/121$.

\section{Analysis and results of optical data}

Using GROND, the counterpart of the {\em Swift} X-ray source was resolved into three point sources.
Star A of \citet{2010ATel.2704....1R} was originally classified as of spectral type B1-2~III from GROND and {\em Swift}/UVOT photometry 
and is the most likely counterpart within the improved \xmm\ X-ray position uncertainty of \xmmp.

The possible northern counterpart at 05\rahour41\ramin26\fs57 $-$69\degr01\arcmin21\farcs7
has a somewhat larger angular separation to the X-ray position of 1.3\arcsec\ (2.6$\sigma$).
The $I$-band light curve derived from OGLE is flat.
GROND photometry ($g^\prime = 17.54 \pm 0.08$,
$r^\prime = 17.55 \pm 0.05$,
$i^\prime = 17.12 \pm 0.05$,
$z^\prime = 16.89 \pm 0.05$,
$J = 16.49 \pm 0.05$,
$H = 16.37 \pm 0.05$, and
$K_s = 16.62 \pm 0.08 $) can be modelled by emission from a K star.
This is too faint and too red for a Be star.
Depending on its luminosity class, the star can be located in the LMC or the Galaxy.
In the latter case, coronal X-ray emission would be significantly softer than observed for \xmmp\ thus we reject this star as possible optical counterpart in the following.
Since there is no soft source detected in the 2001 \xmm\ observation to a limit of 0.005 cts s$^{-1}$, X-ray emission of this star is unlikely contributing to the X-ray spectrum of \xmmp.

\subsection{Photometry}

\begin{figure}
  \resizebox{\hsize}{!}{\includegraphics[angle=0,clip=]{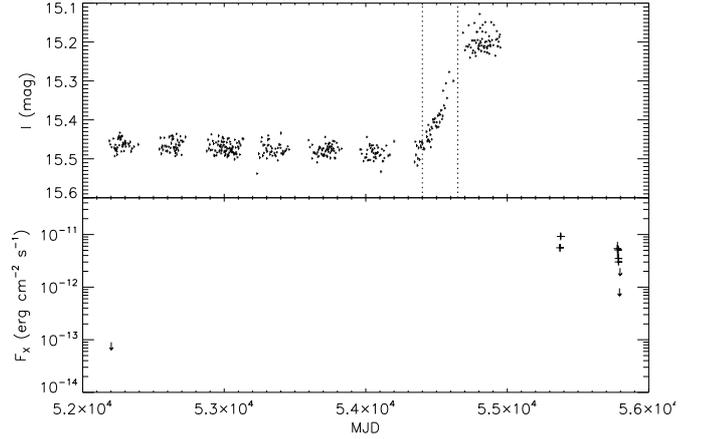}}
  \caption{
           $I$-band light curve of OGLEIII\,LMC175.4.21714 ({\it upper panel}) compared to the X-ray fluxes with upper limits marked by arrows ({\it lower panel}). 
           Dotted lines separate the optical low, transition, and high state. 
          }
  \label{fig:lc}
\end{figure}

The OGLE III $I$-band light curve of this star is presented in Fig.~\ref{fig:lc}.
It exhibits two different brightness states with a transition phase in between.
Before MJD 54400 (left dotted line in Fig.~\ref{fig:lc}), the $I$-band emission is at (15.51--15.43) mag, followed by an impetuous increase to the high state.
After MJD 54650 (right dotted line in Fig.~\ref{fig:lc}), the $I$-band emission is in between (15.24 -- 15.13) mag.

\begin{figure}
  \resizebox{\hsize}{!}{\includegraphics[angle=0,clip=]{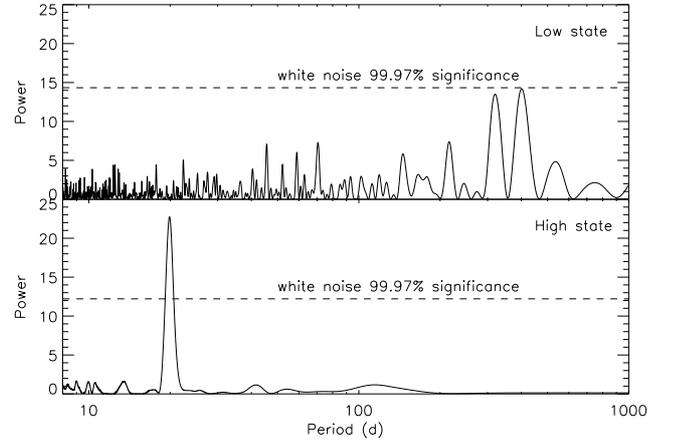}}
  \caption{
           Lomb-Scargle periodogram of the $I$-band of OGLEIII\,LMC175.4.21714 for the low state (upper panel) and high state  (lower panel).
          }
  \label{fig:ls}
\end{figure}

\begin{figure}
  \resizebox{\hsize}{!}{\includegraphics[angle=0,clip=]{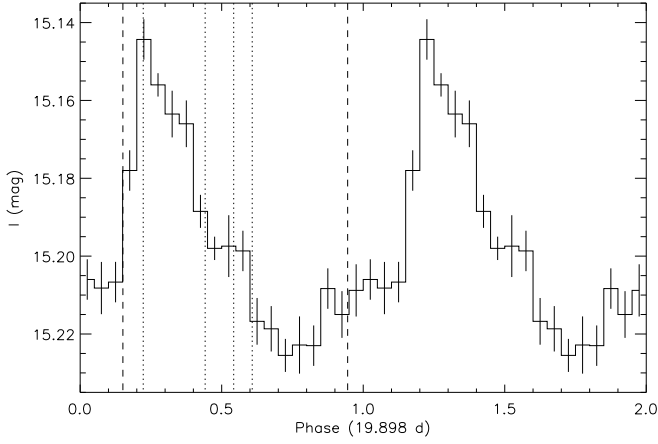}} 
  \caption{$I$-band light curve from OGLE III in the high state convolved with 19.898 days. 
           Phase = 0 corresponds to MJD 54640.
           Dotted lines give the phase of the beginning of the X-ray observations in 2011.
           Dashed lines mark the {\em Swift}/XRT non-detections.
          }
  \label{fig:opt_lc_conv}
\end{figure}

A Lomb-Scargle \citep{1976Ap&SS..39..447L,1982ApJ...263..835S} periodogram of the low and high state of the OGLE III light curve is shown in Fig.~\ref{fig:ls}.
The high state reveals a periodicity with 19.9 days. 
The folded light curve of the high state is presented in Fig.~\ref{fig:opt_lc_conv}.
Dotted and dashed lines mark the phases of X-ray detections and non detections of 2011, respectively.

\begin{table}
\caption{Swift/UVOT photometry.}
\begin{center}
\begin{tabular}{lrr}
\hline\hline
     \multicolumn{1}{l}{Filter} &  
     \multicolumn{1}{c}{UT Date} &
     \multicolumn{1}{c}{AB Mag\tablefootmark{a}}\\
\hline
$v$	&	2010-06-25 06:09	&	15.47	$\pm$	0.06	\\
	&	           07:51	&	15.51	$\pm$	0.06	\\
							
	&	2011-08-10 00:36	&	15.11	$\pm$	0.04	\\
	&	           02:15	&	15.20	$\pm$	0.05	\\
	&	           03:53	&	15.10	$\pm$	0.05	\\
							
	&	2011-08-13 23:07	&	15.13	$\pm$	0.04	\\

$b$	&	2010-06-25 06:02	&	15.36	$\pm$	0.04	\\
	&	           07:45	&	15.43	$\pm$	0.04	\\
	&	           09:28	&	15.34	$\pm$	0.04	\\
	&	2011-08-10 00:24	&	15.08	$\pm$	0.03	\\
	&	           02:06	&	15.08	$\pm$	0.03	\\
	&	           03:48	&	15.07	$\pm$	0.04	\\
	&	2011-08-13 22:55	&	15.07	$\pm$	0.03	\\
	&	2011-08-20 17:13	&	15.10	$\pm$	0.03	\\

$u$	&	2010-06-25 06:00	&	15.41	$\pm$	0.03	\\
	&	           07:43	&	15.38	$\pm$	0.03	\\
	&	           09:26	&	15.35	$\pm$	0.03	\\
							
	&	2011-08-10 00:21	&	15.06	$\pm$	0.03	\\
	&	           02:05	&	15.04	$\pm$	0.03	\\
	&	           03:46	&	15.05	$\pm$	0.03	\\
	&	           05:17	&	15.05	$\pm$	0.05	\\
							
	&	2011-08-13 22:52	&	15.10	$\pm$	0.03	\\
							
	&	2011-08-20 17:11	&	15.12	$\pm$	0.03	\\

$uvw1$	&	2010-06-25 05:58	&	15.45	$\pm$	0.03	\\
	&	           07:41	&	15.44	$\pm$	0.03	\\
	&	           09:24	&	15.44	$\pm$	0.03	\\
							
	&	2011-08-10 00:17	&	15.20	$\pm$	0.03	\\
	&	           02:02	&	15.25	$\pm$	0.03	\\
	&	           03:45	&	15.24	$\pm$	0.03	\\
	&	           05:15	&	15.26	$\pm$	0.03	\\
							
	&	2011-08-13 22:49	&	15.31	$\pm$	0.03	\\
							
	&	2011-08-20 17:08	&	15.22	$\pm$	0.03	\\

$uvm2$	&	2010-06-25 06:19	&	15.56	$\pm$	0.02	\\
	&	           07:59	&	15.57	$\pm$	0.03	\\
							
	&	2010-07-01 00:13	&	15.70	$\pm$	0.03	\\
							
	&	2011-08-10 00:41	&	15.29	$\pm$	0.03	\\
	&	           02:18	&	15.41	$\pm$	0.03	\\
	&	           03:55	&	15.38	$\pm$	0.03	\\
							
	&	2011-08-13 00:31	&	15.33	$\pm$	0.03	\\
	&	           02:06	&	15.33	$\pm$	0.02	\\
	&	           23:12	&	15.43	$\pm$	0.03	\\

$uvw2$	&	2010-06-25 06:05	&	15.60	$\pm$	0.03	\\
	&	           07:48	&	15.63	$\pm$	0.03	\\
	&	           09:30	&	15.61	$\pm$	0.03	\\
							
	&	2010-06-30 01:40	&	15.59	$\pm$	0.02	\\
	&	           03:21	&	15.97	$\pm$	0.02	\\
	&	           04:49	&	15.61	$\pm$	0.02	\\
	&                  17:56	&	15.71	$\pm$	0.02	\\
	&	           19:32	&	15.70	$\pm$	0.02	\\
	&	           22:35	&	15.77	$\pm$	0.03	\\
							
	&	2011-08-10 00:30	&	15.36	$\pm$	0.02	\\
	&	           02:11	&	15.48	$\pm$	0.02	\\
	&	           03:50	&	15.42	$\pm$	0.03	\\
							
	&	2011-08-13 23:01	&	15.54	$\pm$	0.02	\\
							
	&	2011-08-20 01:00	&	15.41	$\pm$	0.02	\\
	&	           17:15	&	15.38	$\pm$	0.04	\\
							
	&	2011-08-24 03:10	&	15.44	$\pm$	0.02	\\
	&	           04:46	&	15.45	$\pm$	0.02	\\

\hline
\end{tabular}
\end{center}
  \tablefoot{\tablefoottext{a}{Not corrected for Galactic foreground reddening or extinction in the LMC.}}
\label{tab:uvot_log}
\end{table}

The {\em Swift}/UVOT photometry is presented in Table~\ref{tab:uvot_log}.
Within uncertainties most values are constant.
For the 2010 measurements, we see an indication for a flux decrease in the $uvm2$ and $uvw2$ magnitudes of $\sim$0.13$\pm$0.04 mag and $\sim$0.17$\pm$0.04 mag, respectively.
Other magnitudes in 2011 are constant, except a possible short increase in the UV on 2011-08-13 between 22:49 and 23:12 by $\sim$0.1~mag.
These are of the same order as the variations observed by OGLE during high and low state.

By comparing the averaged magnitudes of 2010-06-25 and those of 2011, 
we find a stronger flux increase as observed with OGLE III in 2008,
with $\Delta v = -0.36$ mag,
$\Delta b = -0.30$  mag,
$\Delta u = -0.31$  mag,
$\Delta uvw1 = -0.20$  mag,
$\Delta uvm2 = -0.20$  mag, and
$\Delta uvw2 = -0.18$  mag.
This indicates a further transition from low to high state between June 2010 and August 2011.

The GROND magnitudes are summarised in Table~\ref{tab:grond_log}.
Due to pointing constraints,  the first observation was performed only
in the $J$,  $H$, and $K_s$ bands, while the  remaining epochs cover all
seven  channels.
Note the increase in magnitude between 2010-06-26 and 2010-06-29 of
$\Delta J = (0.25\pm0.07)$  mag,
$\Delta H = (0.45\pm0.07)$  mag, and
$\Delta K_s = (0.42\pm0.11)$  mag
within three days.
This affirms the observed drop of the flux as observed with {\em Swift}/UVOT.

\begin{figure}
\centering
\includegraphics[width=0.5\textwidth]{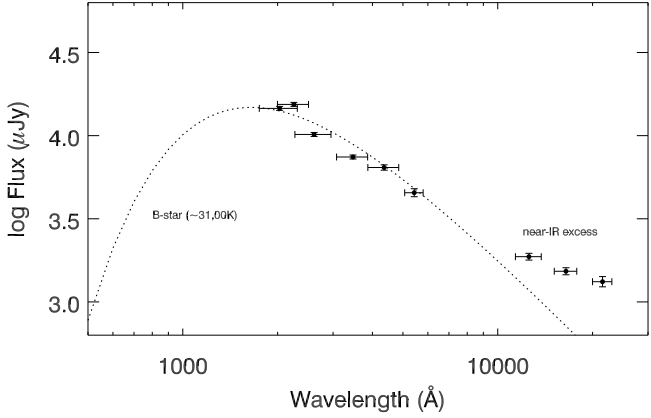}
\caption{UV-near-IR SED composed of {\em Swift}/UVOT observations
  obtained on 2010 June 25th and GROND data taken on 2010 June 26th. 
  The dotted line shows the simplified best-fit black-body model indicating that
  the UVOT photometry is consistent with the B0-1 stellar
  classification derived from the spectroscopy.}
\label{fig:grond+uvotsed}
\end{figure}

The fit to the spectral energy distribution composed
from the first GROND epoch together with averaged UVOT photometry from
2010 June 25th, is shown in Fig.~\ref{fig:grond+uvotsed}. Here, all magnitudes were
corrected for Galactic reddening of E$_{\rm B-V}=0.075$\,mag
\citep{1998ApJ...500..525S} using the \citet{1989ApJ...345..245C} extinction
law and for the LMC-intrinsic reddening of E$_{\rm B-V}=0.15$ mag
(see Sec.~\ref{specclass}) using the \citet{1992ApJ...395..130P} extinction law. The data are
best fit with a hot ($\approx31\,000$ K) black-body spectrum, consistent
with the B0-1 III stellar classification suggested by the optical
spectroscopy (Sec.~\ref{specclass}). 
We note the existence of a clear excess in the near-IR bands.

\begin{table*}
\caption{GROND photometry.}
\begin{tabular}{lccccccc}
\hline\hline
UT Date&\multicolumn{7}{c}{AB Magnitude\tablefootmark{a}} \\
& $g^\prime$ & $r^\prime$ & $i^\prime$ & $z^\prime$ & $J$ & $H$ & $K_s$\\
\hline
2010-06-26 10:51 & -- & -- & -- & -- & $15.93\pm0.05$ & $16.07\pm0.05$ & $16.29\pm0.07$ \\
2010-06-29 10:29 & $15.34\pm0.04$ & $15.60\pm0.03$ & $15.82\pm0.03$ & $15.98\pm0.04$ & $16.17\pm0.05$ & $16.53\pm0.05$ & $16.71\pm0.08$  \\
2012-01-25 04:14 & $15.18\pm0.04$ & $15.41\pm0.03$ & $15.57\pm0.03$
& $15.72\pm0.04$ & $15.96\pm0.05$ & $16.28\pm0.05$ & $16.54\pm0.08$ \\
\hline
\end{tabular}
\tablefoot{\tablefoottext{a}{Not corrected for Galactic foreground reddening or extinction in the LMC.}}
\label{tab:grond_log}
\end{table*}

\subsection{Spectral classification}
\label{specclass}

OB stars in our own galaxy are classified using the ratio of certain metal and helium lines \citep{1990PASP..102..379W} 
based on the Morgan-Keenan \citep[MK; ][]{1943QB881.M6.......} system. 
However, this is unsuitable in lower metallicity environments as the metal lines are either much weaker or not present. 
As such, the optical spectrum of IGR~J05414-6858 was classified using the method developed by \citet{1997A&A...317..871L} for B-type stars in the SMC 
and implemented for the SMC, LMC and Galaxy by \citep{2004MNRAS.353..601E,2006A&A...456..623E}. 
This system is normalized to the MK system such that stars in both systems show the same trends in their line strengths. 
The luminosity classification method from \citet{1990PASP..102..379W} was assumed in this work.

\begin{figure*}
\centering
 \includegraphics[height=180mm,angle=90]{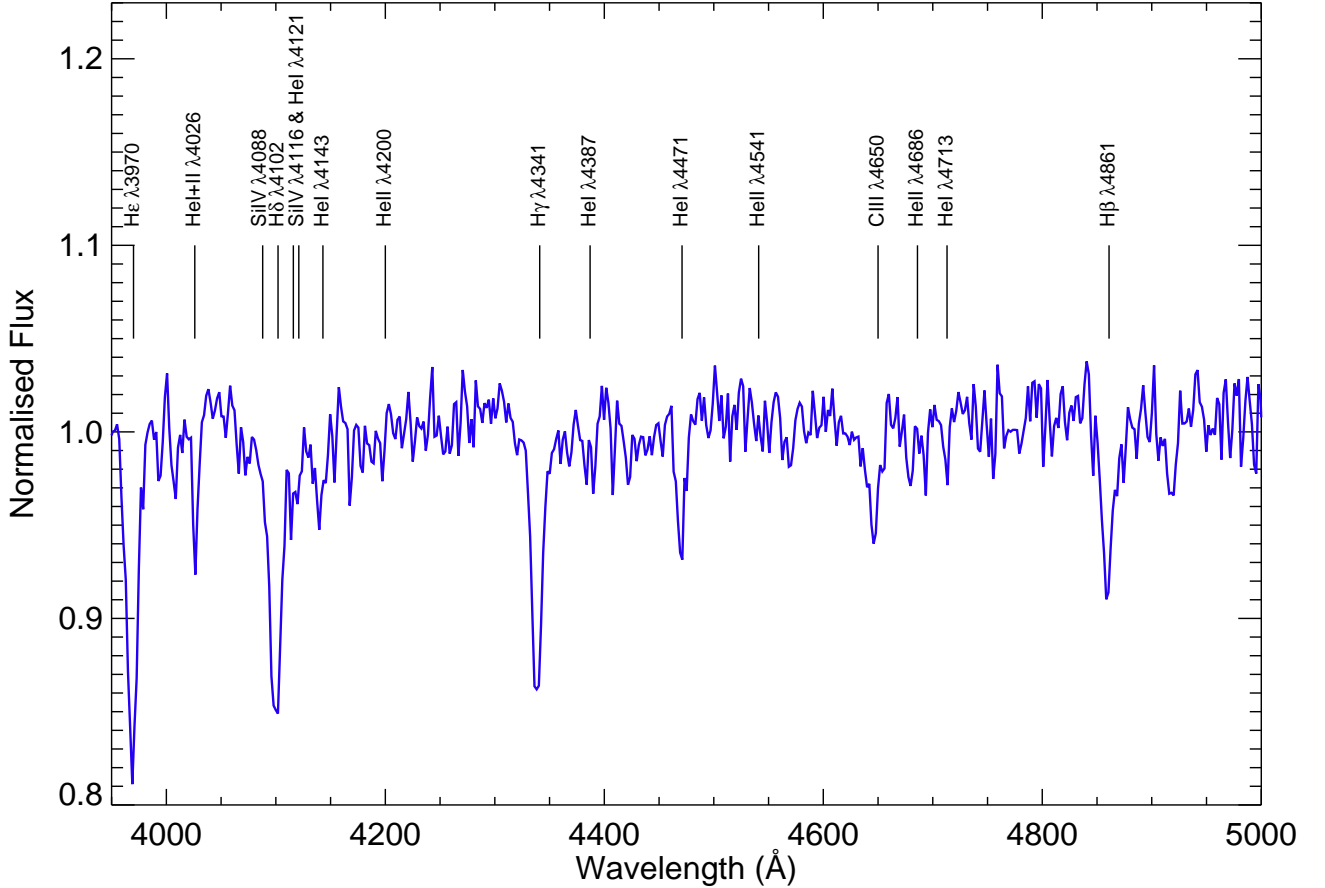}
\caption{Spectrum of IGR~J04514-6858 in the wavelength range $\lambda\lambda$3900--5000\AA{} with the NTT on 2011-12-08. 
         The spectrum has been normalized to remove the continuum and redshift corrected by -280~km~s$^{-1}$. 
         Atomic transitions relevant to spectral classification have been marked.}
\label{fig:blue}
\end{figure*}

Fig.~\ref{fig:blue} shows the unsmoothed optical spectrum of IGR~J04514-6858. 
The spectrum is dominated by the hydrogen Balmer series and neutral helium lines. 
The \ion{He}{i} line at $\lambda4143$~\AA{} is stronger than the \ion{He}{ii} $\lambda4200$\AA{}, which means the star is later than type O9. 
\ion{He}{ii} $\lambda4686$~\AA{} is also clearly present implying that the optical counterpart of IGR~J04514-6858 is earlier than type B1.5, 
although the He\textsc{ii} $\lambda4541$~\AA{} line is not visible above the noise level of the data. 
There is also evidence for the \ion{Si}{iv} $\lambda4116$~\AA{} line - consistent with a spectral classification of B1. 
However \citet{1990PASP..102..379W} present spectra of B0 type stars with clear Si lines indicating that a spectral classification of B0 is not ruled out by their presence. 
We note that there does not appear to be any evidence for the \ion{Si}{iv} $\lambda4088$~\AA{} line. 
Be stars are characterised by their rapid rotation velocity and it could be that this line is being concealed by the rotationally broadened H$\delta$ line in close proximity.

\begin{figure}
\centering
 \includegraphics[height=90mm,angle=90]{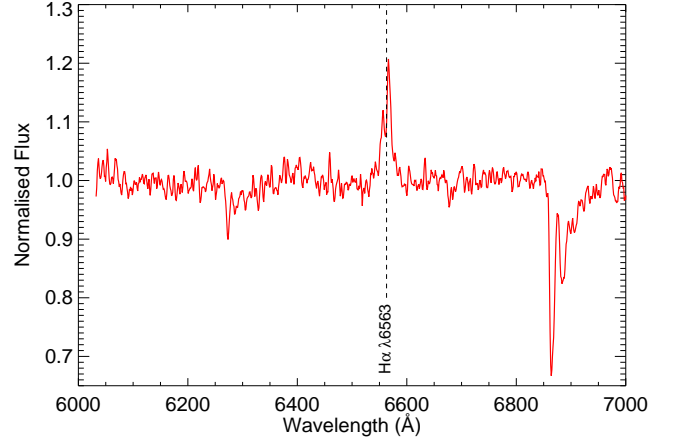}
\caption{Spectrum of IGR~J04514-6858 in the wavelength range $\lambda\lambda$6000--7000\AA{} with the NTT on 2011-12-10. 
         The spectrum has been smoothed with a boxcar average of 3, normalized to remove the continuum and shifted by -280~km~s$^{-1}$.}
\label{fig:red}
\end{figure}

The luminosity class of the system was determined using the ratios of \ion{S}{iv} $\lambda4116$/\ion{He}{i} $\lambda4121$, \ion{He}{i} $\lambda4121$/\ion{He}{i} $\lambda4143$ and \ion{He}{ii} $\lambda4686$/\ion{He}{i} $\lambda4713$. 
The first two ratios strengthen with decreasing luminosity class (i.e. with increasing luminosity), whereas the latter ratio decreases with increasing luminosity. 
The relative strengths of these lines suggest a luminosity class III, making our spectral classification of B0-1~III consistent with that obtained photometrically with the GROND and {\em Swift}/UVOT data. 
To check the spectral identification we can compare the observed optical magnitudes with that predicted for a B0-1 III star in the LMC. 
Taking the faintest, least disc-contaminated $V$-band magnitude from Table~\ref{tab:uvot_log} of $V=(15.51\pm0.06)$ mag, 
a distance modulus of 18.5$\pm$0.1 \citep{2009AJ....138....1K} and a reddening of E$_{\rm B-V}=0.15$ mag \citep{1991A&A...246..231S} reveals an absolute magnitude for the star of $M_V= (-3.5\pm0.2)$.
Such a value would be consistent with a B0.5 III star \citep{2006MNRAS.371..185W} and hence confirms the classification deduced from the spectra reported here. 
We note that we cannot use the derived X-ray absorption reported here to refine the column to the star because of the large uncertainties in that value.

Fig.~\ref{fig:red} shows the red end of the spectrum of IGR~J04514-6858 taken near-simultaneously. The H$\alpha$ equivalent width, considered an indicator for circumstellar disc size, 
is relatively small at $-(3.2\pm0.6)$\AA{} 
which is consistent with the lack of H$\beta$ in emission in Fig.~\ref{fig:blue}. 
The double peaked, asymmetric nature of the line profile shows a \emph{V/R} pattern consistent with global one armed oscillations (GOAO) and suggests that the circumstellar disc of the star is inclined to the line of sight. This is not uncommon in the circumstellar discs of Be stars.

\section{Discussion and conclusions}
\label{sec:discussion}

We performed an \xmm\ ToO observation of \xmmp\ in August 2011, allowing us to measure the X-ray spectrum and discover the spin period.
This adds the tenth known HMXB pulsar to the LMC sample and confirms the neutron star nature of the compact object.

The source was found in INTEGRAL observations performed on 2010 May 13--22 (130 ks) and June 6--14 (400 ks) 
at an average flux of $8\times10^{36}$ erg s$^{-1}$ in the (20--40) keV band \citep{2010ATel.2695....1G}.
In the Swift follow-up observation on June 30 we see the source still in outburst at a luminosity of $\sim$$3\times10^{36}$ erg s$^{-1}$ in the (0.2--10.0) keV band.
If these detections correspond to the same outburst, the duration of X-ray bright state would suggest a type II outburst.
The luminosity, derived from INTEGRAL is at the lower limit for a classical type II outburst, but might have been higher at maximum.
The luminosity and observed duration of the 2011 outburst are in agreement with a type I outburst, 
but since the time of the beginning of the outburst is unknown, we cannot exclude a type II outburst.

Furthermore, the spectrum of the current outburst was found to be significantly harder than in 2010.
The power-law photon index of the 2011 outburst of $\Gamma$ = 0.3--0.4 is also relatively low, compared to the distribution known from the SMC sample ($\Gamma\sim1$). 
We note that also for a few BeXRBs in the SMC hard spectra were observed, as for the pulsars 
XTE\,J0103-728 \citep[$\Gamma=0.35-0.54$, $P_{\rm spin} = 6.85$ s, ][]{2008A&A...484..451H} and
XMMU\,J004814.0-732204 \citep[$\Gamma=0.53-0.66$, $P_{\rm spin} = 11.87$ s, ][]{2011A&A...527A.131S}.
The first one of these also was detected in a type II outburst with INTEGRAL \citep{2010MNRAS.403.1239T}. 
Moreover, for both SMC pulsars an indication of a soft excess with comparable emission radius was found
and suggested to originate from the accretion disc.
The system intrinsic absorption strongly depends on the modelling ($0-5\times 10^{21}$ cm$^{-2}$) and the location of the system in the LMC.
The total LMC column density along the line of sight at the position of \xmmp\ is $3.6\times 10^{21}$ cm$^{-2}$ \citep{2003ApJS..148..473K}.

According to the Corbet relation \citep{1984A&A...141...91C,2005ApJS..161...96L,2009IAUS..256..361C},
we expect the orbital period of the system in the range of (1 -- 100) days for the measured spin period of \xmmp.
The periodic variations seen in the $I$-band are therefore likely caused by binarity.
Assuming masses of 21.5\msun\ and 1 \msun\ for the Be star and the NS, respectively, the orbital period implies a semi-major axis of the 
binary system of 0.41 AU, corresponding to $\sim$6 stellar radii \citep{1996ApJ...460..914V}. 
The X-ray detections in 2011 occurred during the bright phase of the folded $I$-band light curve (see dotted lines in Fig.~\ref{fig:opt_lc_conv}).
In 2011, the X-ray luminosity follows the $I$-band emission and the non-detections were during low $I$-band emission.
A correlation of X-ray and optical outbursts was e.g. reported for AX J0058-720 \citep{2007A&A...476..317H}.
In the case of the 2010 outburst, the two {\em Swift} X-ray detections are during low $I$-band emission at phase 0.80 and 0.04.
This is either due to the fact, that type II outbursts are not correlated to the orbital phase,
or that the optical period is not caused by binarity.

Long-term optical variation with different variability patterns are typical for BeXRBs \citep{2011MNRAS.413.1600R}.
An explanation for the transition from low to high state by $\Delta I \sim 0.3$ mag might be the build up of a decretion disc around the Be star.
The \xmm\ upper limit (X-ray faint state) in 2001 was during the optical low state, where probably no decretion disc was present.
From other BeXRBs \citep[c.f Fig. 10 of ][]{2011Ap&SS.332....1R}, a correlation between NIR and optical magnitudes is seen.
Therefore it is likely that the long term variability seen in the OGLE $I$-band extends from NIR to UV.
As indicated by the GROND and {\em Swift}/UVOT observations, the system still undergoes strong variations in the NIR, optical and UV.
A rapid strong drop in the NIR emission was observed between June 2010 26th and 29th. 
At this time, the source was still in a presumable type II X-ray outburst, 
which can be followed by a disc loss phase.
The NIR-flux decrease further supports a type II outburst.
Also, the $K_s$-band magnitude was high, compared to the GROND January 2012 observation, 
while $J$ was at the same level again pointing to the presence of a circumstellar disc.
This is further supported by the H$\alpha$ line emission in December 2011.
Unfortunately, the end of the 2010 X-ray outburst is not constrained.
During the 2011 outburst, the {\em Swift}/UVOT observations do not indicate any strong changes in the optical.
Forthcoming OGLE IV data will allow to extend the light curve and to confirm the periodicity.

IGR\,J05414-6858 is the tenth known HMXB pulsar in the LMC with $P_{\rm spin} = 4.4208$ s.
The optical counterpart was classified as of spectral type B0-1~IIIe and shows double-peaked H$\alpha$ emission and a variable NIR excess.
A likely orbital period of $P_{\rm orb} = 19.9$ d was found.
The two observed X-ray outbursts demonstrate the importance of optical monitoring during outbursts, to better understand the accretion process in these systems.
To further increase the sample of explored HMXB in the LMC, further X-ray observations triggered during an outburst are necessary.

\begin{acknowledgements}
We thank the \xmm\ team for scheduling the ToO observation.
The XMM-Newton project is supported by the Bundesministerium f\"ur Wirtschaft und 
Technologie/Deutsches Zentrum f\"ur Luft- und Raumfahrt (BMWI/DLR, FKZ 50 OX 0001)
and the Max-Planck Society. 
Part of the funding for GROND (both hardware
as well as personnel) was generously granted from the Leibniz-Prize to
Prof.   G.   Hasinger  (DFG  grant  HA~1850/28-1).
The OGLE project has received funding from the European Research Council
under the European Community's Seventh Framework Programme
(FP7/2007-2013) / ERC grant agreement no. 246678 to AU.
We acknowledge the use of public data from the {\em Swift} data archive.
X-LZ acknowledges financial support by DLR FKZ 50 OG 0502. 
RS acknowledges support from the BMWI/DLR grant FKZ 50 OR 0907.
\end{acknowledgements}

\bibliographystyle{aa}
\bibliography{../auto,../general}

\end{document}